\documentclass[prd,aps,twocolumn,nofootinbib,fleqn]{revtex4}
\usepackage{graphicx}
\usepackage[colorlinks]{hyperref}
\usepackage{xspace}
\usepackage{mathtools}
\usepackage{amsmath}
\usepackage{gensymb}

\def\xmm{\textit{XMM-Newton}\xspace}
\def\theseus{\textit{THESEUS}\xspace}

\begin{document}

\title{THESEUS Insights into ALP, Dark Photon and Sterile Neutrino Dark Matter}

\author{Charles Thorpe-Morgan$^1$, Denys Malyshev$^1$, Andrea Santangelo$^1$, Josef Jochum$^2$, Barbara J\"ager$^2$, Manami Sasaki$^3$, Sara Saeedi$^3$}
\address{$^{1}$ Institut f{\"u}r Astronomie und Astrophysik T{\"u}bingen, Universit{\"a}t T{\"u}bingen, Sand 1, D-72076 T{\"u}bingen, Germany \\
$^2$ Eberhard Karls Universit{\"a}t T{\"u}bingen, 72076 T{\"u}bingen, Germany \\
$^3$ Remeis Observatory and ECAP, Universit{\"a}t Erlangen-N{\"u}rnberg, Sternwartstrasse 7, 96049 Bamberg, Germany
}

\begin{abstract}

Through a series of simulated observations, we investigate the capability of the instruments aboard the forthcoming \theseus{} mission for the detection of a characteristic signal from  decaying dark matter (DM) in the keV-MeV energy range. We focus our studies on three well studied Standard Model extensions hosting axion-like particle, dark photon, and sterile neutrino DM candidates.
We show that, due to the sensitivity of \theseus{}'  X and Gamma Imaging Spectrometer (XGIS) instrument, existing constraints on dark matter parameters can  be improved by a factor of up to  $\sim{}300$, depending on the considered DM model and assuming a zero level of systematic uncertainty. We also show that even a  minimal level of systematic uncertainty of $1\%$ can impair potential constraints by one to two orders of magnitude. We argue that nonetheless, the constraints imposed by \theseus{} will be substantially better than existing ones and will well complement the constraints of upcoming missions such as \textit{eXTP} and \textit{Athena}. Ultimately, the limits imposed by \theseus{} and future missions will ensure a robust and thorough coverage of the parameter space for decaying DM models, enabling either a detection of dark matter or a significant improvement of relevant limits.

\end{abstract}

\maketitle

\section{Introduction}
\label{sec:intro}
Dark matter (DM) remains to be one of the greatest obstacles to our understanding of cosmology. The presence of a universally pervading extra mass is clear and has been precisely measured ($\Omega_{DM} = 0.2641\pm0.0002$~\cite{planck2019}); however, apart from its presence in the Universe, the nature and properties of dark matter remain elusive. The Standard Model (SM) is now known not to host any viable dark matter candidate particles, which has led to the consideration of various extensions to the SM that host potential dark matter candidates,  see~\cite{review_old,pospelov08,sterile_neutrino_review19} for recent reviews. 

A very general low energy extension of the SM is comprised of a ``dark sector'', so called due to its extremely weak interaction with the SM. While such a sector can, in principle, host a variety of new particles providing natural DM candidates (see~\cite{battaglieri17} for a review) and self-interactions, it can most easily be accessed via interactions between the dark and SM sectors. Such cross-sector interactions are often undertaken through a ``mediator'' -- a particle with both, SM and dark sector, quantum numbers. Alternatively the SM particles can  interact with the DM particles either directly (if they possess charge under the corresponding interaction) or through mixing. Some representative types of DM models are~\cite{essig13,raggi15}:\\
-- models with (pseudo)scalar DM particles, e.g. axions and axion-like particles (ALPs);\\
-- models with sterile neutrinos acting as DM particles; \\
-- models with a vector DM particle (e.g. a dark photon (DP)).

In the following, we investigate three well studied cases of these models with massive DM candidates, which have the potential to comprise the majority of the observed dark matter. 
Namely, we will consider ALPs, sterile neutrinos and dark photons as dark matter candidates.

In these models a dark matter particle can decay, consequentially emitting photons. An axion or ALP $a$ can decay into two photons $a\rightarrow\gamma+\gamma$. On the other hand, sterile neutrino dark matter $N$ can manifest itself via a two body decay: $N\rightarrow \nu+\gamma$, while dark photons $V$ are subject to three-photon decay $V\rightarrow 3\gamma$ (a preferable decay channel for $m_V < m_e$~\cite{pospelov08}).

The foremost consequence of such radiative decay channels would be the potential for a detectable signal originating from DM dominated astrophysical objects. The detection of such a signal would allow the indirect detection of dark matter decay events. 

While, generally, any astrophysical object with a high DM concentration can be a target for such searches, one must consider additional astrophysical properties of the object to analyse its feasibility as the focus of such a search. For example, the high DM density and a low level of astrophysical background makes dwarf spheroidal galaxies (dSphs) advantageous targets for dark matter searches across a considerable section of the electromagnetic spectrum. However, dSphs are at most degree-scale targets, and such a small angular size does not allow for the full utilisation of the capabilities of broad field of view (FoV) instruments. Conversely, when considering wide FoV instruments, much wider objects with angular extensions of close to the whole sky (e.g. DM halo of the Milky Way) are preferable targets for these instruments.

Despite numerous searches, no clear evidence for any of the described dark matter candidates has been found so far. These searches have however allowed for parameters (mass and/or coupling strengths) to be significantly constrained  for all candidates considered in this work. 

The fundamental limit on the sterile neutrino mass $m_N\gtrsim 1$~keV arises from the requirement that the phase space density of DM particles in the halos of dSphs may not exceed the limits imposed by the uncertainty relation and the initial phase  space density at the moment of production of DM in the Early Universe~\cite{tremainegunn,tg1,gorbunov08,savchenko19}. The mixing with active neutrinos is also constrained from above and below by the non-detection of a decay line in astrophysical observations and the exclusion of values that would lead to a discrepancy between observed and predicted abundances of light elements produced during Big Bang Nucleosynthesis~\cite{shi99,serpico05,shaposhnikov08,laine08,canetti13,kusenko06}. 

The best limits on ALPs in different energy bands are based on observations of objects of a totally different nature. These include: astrophysical observations (non-detection of a decay line in the $\gamma$-ray background) in the keV-MeV mass range; limits based on the evolution of horizontal branch stars (eV-keV masses); or direct detection experiment limits and astrophysical limits based on non-detection of ALP-photon conversion in certain magnetised astrophysical objects, see~\cite{pospelov08,graham13} for  reviews.

The parameters of Dark photons are subject to constraints from the non-observation of a characteristic feature in the spectrum of the galactic diffuse background (for masses $m_V\gtrsim 10$~keV);  stellar-evolution constraints (including the Sun, horizontal branch, and red giant stars~\cite{redondo13}) for masses $m_V\sim 10^{-6} - 10^4$~eV; cosmological and direct detection experiment limits at lower masses, see~\cite{pospelov08,fabbrichesi20} for reviews.

In what follows we study the capabilities of the forthcoming Transient High Energy Sky and Early Universe Surveyor (\theseus{}~\cite{theseus_1,theseus_2,stratta2018theseus,tanvir2018theseus} ) mission to constrain parameters of keV-MeV mass scale dark matter focusing on the candidates described above.

\theseus is a European mission concept\footnote{selected by ESA on 2018 May 7 to enter an assessment phase study.} designed in response to the ESA call for medium-size mission (M5) within the Cosmic Vision Program\footnote{\href{https://www.esa.int/Science_Exploration/Space_Science/ESA\_s\_Cosmic\_Vision}{See Cosmic Vision Program website}}. The fundamental goals of the \theseus{} mission are the study and detection of high energy transient phenomena, the study of the early universe and the epoch of re-ionisation, and ``the hot and energetic universe". These goals are planned to be achieved using the mission's unique combination of instruments.

 The \theseus{} mission will host a total of three telescope arrays, covering a section of the infrared regime as well as the energy range of soft and hard X-rays. The proposed instrumental payload for \theseus{} is:\\
 --The Soft X-Ray Imager (SXI), an array of 4 lobster-eye~\citep{angel97} telescope units with a quasi-square FoV covering the energy range of $0.3 - 5$~keV with an effective area of $A_{eff}\approx 1.9$~cm$^2$ at 1 keV and an energy resolution $\sim 4$\%. These will cover a total FoV of $\sim 1$~sr with source location accuracy $<1-2$ arcminutes (for a full review of the instrument see \cite{SXI}).\\
 --The InfraRed Telescope (IRT), a single large (0.7~m) telescope that will be used for follow-up observations of gamma-ray bursts. It will operate in the wavelength band  $0.7 - 1.8\, \mu$m and have a $15'\times 15'$ FoV (for further specifications on the IRT see \cite{IRT}).\\
 --The X-Gamma Ray Imaging Spectrometer (XGIS) array, consisting of coded-mask cameras (with the total half-sensitive FoV comparable to that of the SXI)  using monolithic X-gamma ray detectors based on bars of silicon diodes coupled with CsI crystal scintillator. XGIS will operate in the energy range of 2~keV -- 20~MeV, which will be achieved using the two different detectors, referenced hereafter as XGIS-X and XGIS-S. The Silicon Drift Detector (SDD) will cover the energy range of 2--30~keV (XGIS-X) whereas the CsI scintillator will cover the range of 20~keV -- 2~MeV (XGIS-S\footnote{Note, that due to the transparency of the XGIS coded mask at hard X-rays at $E\gtrsim 150$~keV XGIS-X operates as a collimator.}). The effective areas and energy resolutions of XGIS-S are $A_{eff}(300\,\mbox{keV})\approx 1100$~cm$^2$ and energy resolution changing from $\Delta E/E\sim 15$\% at below $100$~keV to $\Delta E/E\sim 2$\% at higher energies. The effective area and resolution of XGIS-X instrument are $A_{eff}(10\,\mbox{keV})\approx 500$~cm$^2$ and $\Delta E/E\sim 1.5$\%, see \cite{XGIS} for the full technical proposal for the XGIS. We summarise these technical characteristics and compare them to current and next-generation missions in Tab.~\ref{tab:missions}.

\begin{figure*}
\includegraphics[width=0.48\linewidth]{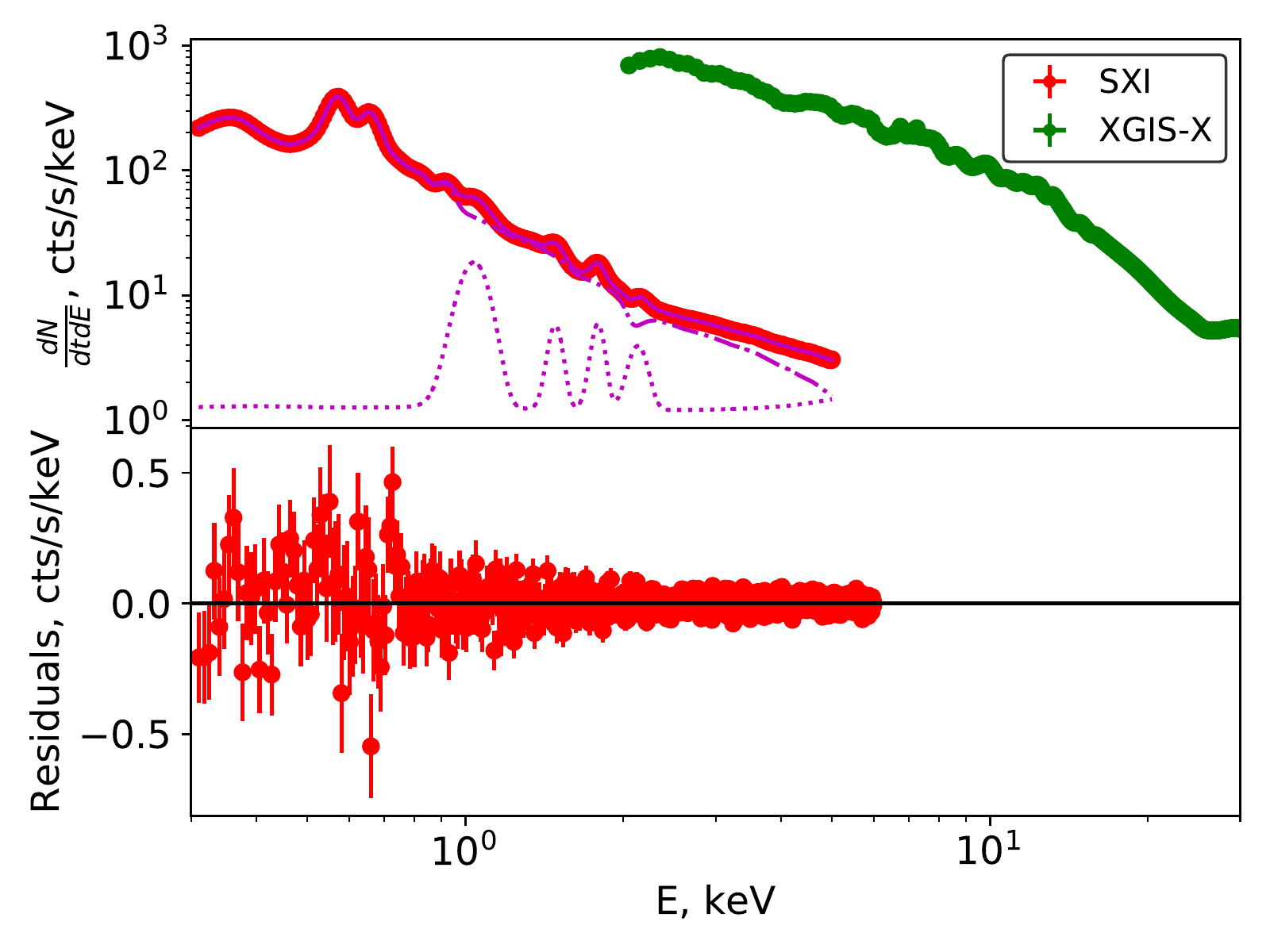}
\includegraphics[width=0.48\linewidth]{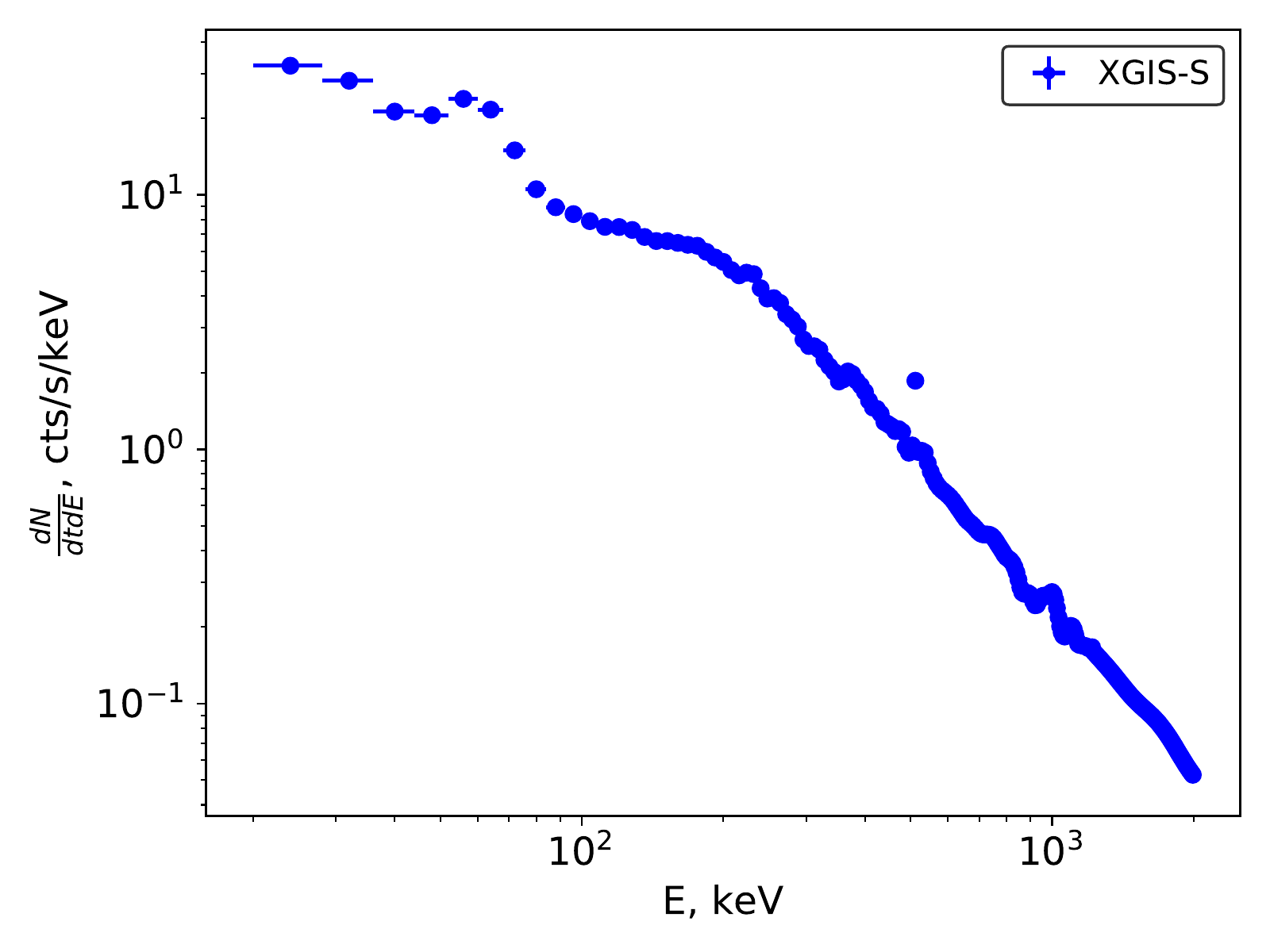}
   \caption{Spectra of blank sky observations from \theseus' instruments: the SXI, XGIS-X (left panel) and XGIS-S  (right panel). The lower sub-panel  illustrates the residuals (model subtracted from data) of the SXI instrument according to the best-fit model (solid magenta line). Dotted and dot-dashed magenta lines represent the instrumental and sky components of the model, see text for more details.}
    \label{fig:theseus_spec}
\end{figure*}

 Focusing on keV-MeV mass scale dark matter, we omitted the IRT from our further investigations. However, both the SXI and the XGIS have a large potential for the detection of DM decay given their very large FoVs (see e.g. discussion in~\cite{zhong20}), thus the sensitivity simulations run by this study were performed for both these instruments.

\begin{table}[t]
    \centering
    \begin{tabular}{|c|c|c|c|c|c|}
    \hline
        Instrument & $A_{eff}^{peak}$ & $E^{peak}$ & $dE/E^{peak}$ & FoV & Date\\
         & cm$^2$ & keV &  & sr & \\
        \hline
        SXI &1.9&1&0.04&1& 2032\\
        XGIS-X &504&8.5&0.06&1& 2032\\
        XGIS-S &1060&350&0.024&1& 2032\\
        \hline
        Athena/X-IFU & $1.6\cdot 10^4$ & 1.4 & $1.9\cdot 10^{-3}$ & $3.3\cdot 10^{-6}$ & 2031 \\
        eXTP/SFA &8600&1.5&0.1&$9.6\cdot 10^{-6}$& 2027\\
        eXTP/WFM &77&9&0.029&2.5& 2027\\
        \hline
        Hitomi/SXS & 84 & 6 & $9.4\cdot 10^{-4}$ & $7.1\cdot 10^{-7}$ &2016-2016\\
        XMM/PN &815&1.5&0.07&$4.5\cdot 10^{-5}$& 1999-**\\
      \hline  
    \end{tabular}
    \caption{
    Technical characteristics of THESEUS
    compared to current and next generation missions.
    The table summarises the peak effective area Apeak, its
    corresponding energy Epeak and the energy resolution
    at this energy dE=Epeak as well as the field of view
    of the instrument. The ''Date" column summarises
    either: the planned launch date of future missions (as
    of 2021), or the launch and de-orbit date of current
    generation missions.
    We note that the parameters of the Hitomi/SXS are similar to those of the XRISM mission, planned for 2022~\citep{xrism}.} 
    \label{tab:missions}
\end{table}

Following this introduction, we present the methodology of our study of the capabilities of the forthcoming \theseus{} mission to probe the parameter space of DM models with ALPs, sterile neutrinos and dark photons.

\section{Search for decaying DM with \theseus{}}
\label{sec:dm_decay_search}

The decay of massive ($m_{DM}$) DM particles with an emission of $\mu$ photons in each decay will result in the photon spectrum (as a function of photon energy $E$)
\begin{align}
\label{eq:dm_spectrum_generic}
& \frac{d\mathcal{F}}{d\Omega}\equiv \frac{dN}{dEdtdAd\Omega} =\frac{1}{4\pi}\cdot\frac{\mu D}{m_{DM}}\cdot \frac{d\Gamma(E)}{dE}    
\end{align}
and a corresponding spectrum in the total field of view of the observing instrument:
\begin{align}
\label{eq:dm_flux_generic}
& \mathcal{F}_{FoV} =\int\frac{d\mathcal{F}}{d\Omega}d\Omega = \frac{\mu D_{FoV}}{4\pi m_{DM}}\cdot \frac{d\Gamma(E)}{dE}\,.
\end{align}

 The $D_{FoV}$ term in Eq.~(\ref{eq:dm_flux_generic}) is the total D-factor (DM mass column density) within the field of view and represents the astrophysical component of the dark matter signal. This is defined as the integral of the DM density over the field of view of the instrument and the line of sight ($l.o.s$), i.e.
\begin{align}
\label{eq:D-factor}
& D_{FoV} \equiv \int D d\Omega = \int\limits_{FoV}\int\limits_{l.o.s.}\rho_{DM}d\ell d\Omega \,.
\end{align}

The  $\Gamma$ term in Eqs.~(\ref{eq:dm_spectrum_generic})-(\ref{eq:dm_flux_generic}) represents the radiative decay width -- a model-dependent term which for the three differing models considered in this study is described below. $\Gamma$ for the $\nu MSM$ (sterile neutrino) model, is given by~\cite{pal,barger95}:
\begin{align}
\label{eq:gamma_MSM}
& \left. \frac{d\Gamma}{dE}\right|_{\nu MSM} = \frac{9\alpha G^2_F}{256 \cdot 4\pi^4}\sin^2(2\theta)m^5_{N}\delta(E-m_{N}/2) 
\end{align}

Here $m_N$ is the mass of the sterile neutrino; $\theta$ is the mixing angle and $\alpha_{QED}$ and $G_F$ stand for fine structure and Fermi constants. 

For Axion Like Particles, $d\Gamma/dE$ is of the form~\citep{pospelov08, pdg2019}
\begin{align}
\label{eq:gamma_ALP}
&\left. \frac{d\Gamma}{dE}\right|_{ALP} = \frac{g_{a\gamma\gamma}^2m_{a}^3}{64\pi\hbar}\delta(E-m_{a}/2) 
\end{align}

In this equation $m_a$ denotes the mass the ALP; whereas $g_{a\gamma\gamma}$ represents the ALP-photon coupling strength. 

Finally the value of $\Gamma$ for Dark Photons is given by the equation~\citep{pospelov08, an15},

\begin{align}
\label{eq:gamma_DP}
& \left. \frac{d\Gamma}{dE}\right|_{DP} = \frac{2\kappa^2\alpha^4_{QED}}{2^73^75^3\pi^3\hbar}\left(\frac{m_{V}}{m_e}\right)^8\cdot f(x); \\ \nonumber
& f(x)=x(1715-3105x+\frac{2919}{2}x^2) ; \\ \nonumber
& x\equiv\frac{2E}{m_{V}};x\in[0;1]\,,\\ \nonumber
\end{align}

where again $m_V$ is the mass of the dark photon, and $\kappa$ is the DP kinetic mixing parameter.

Substituting the respective expressions of Eqs.~(\ref{eq:gamma_MSM} -- \ref{eq:gamma_DP}) into Eq.~(\ref{eq:dm_flux_generic})  and utilising values from~\cite{pdg2019}, 
one obtains the form of expected signal for each model, 

\begin{align}
\label{eq:dm_flux}
& \mathcal{F}_{\nu MSM}(E)\approx 10^{-7} \left(\frac{\sin^2(2\theta)}{10^{-16}}\right)\left(\frac{m_{N}}{10\,\mbox{keV}}\right)^4\times \\ \nonumber
& \times\left(\frac{D_{FoV}}{10^{22}\,\mbox{GeV/cm}^2}\right)\delta(E-m_N/2)\frac{\mbox{ph}}{\mbox{cm}^2\mbox{s\,keV}}; \\ \nonumber
& \mathcal{F}_{ALP}(E)\approx 1.2\cdot 10^{-7}\left(\frac{m_a}{10\,\mbox{keV}}\right)^2\left(\frac{g_{a\gamma\gamma}}{10^{-20}\,\mbox{GeV}^{-1}}\right)^2 \\ \nonumber
&\times\left(\frac{D_{FoV}}{10^{22}\,\mbox{GeV/cm}^2}\right)\delta(E-m_a/2)\frac{\mbox{ph}}{\mbox{cm}^2\mbox{s\,keV}}; \\ \nonumber
&  \mathcal{F}_{DP}(E)\approx 4.08\cdot10^{-7}\left(\frac{\kappa}{10^{-10}}\right)^2\left(\frac{m_V}{10\,\mbox{keV}}\right)^7 \\ \nonumber
&\times f\left(\frac{2E}{m_V}\right) \left(\frac{D_{FoV}}{10^{22}\,\mbox{GeV/cm}^2}\right)\frac{\mbox{ph}}{\mbox{cm}^2\mbox{s\,keV}};
\end{align}

Here we adopted values of the known fundamental constants (e.g $\alpha$, $G_F$, etc.) from~\cite{pdg2019}, we also scaled parameters for some characteristic values and finally accounted for the production of $\mu=1$ photon in sterile neutrino decays, $\mu=2$ photons for ALP decays and $\mu=3$ for three-photon dark photon decay.
\begin{figure*}
\includegraphics[width=0.48\linewidth]{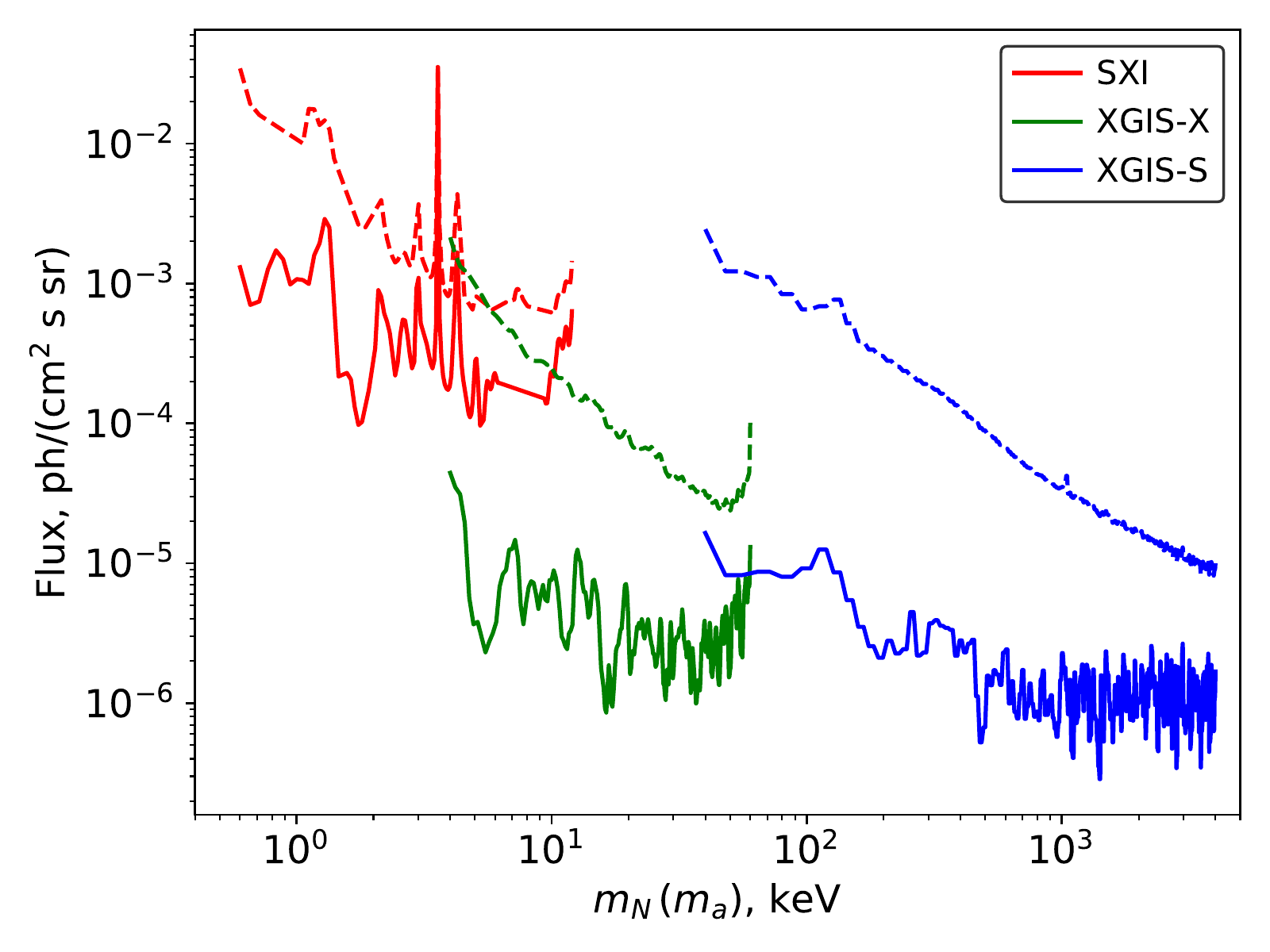}
\includegraphics[width=0.48\linewidth]{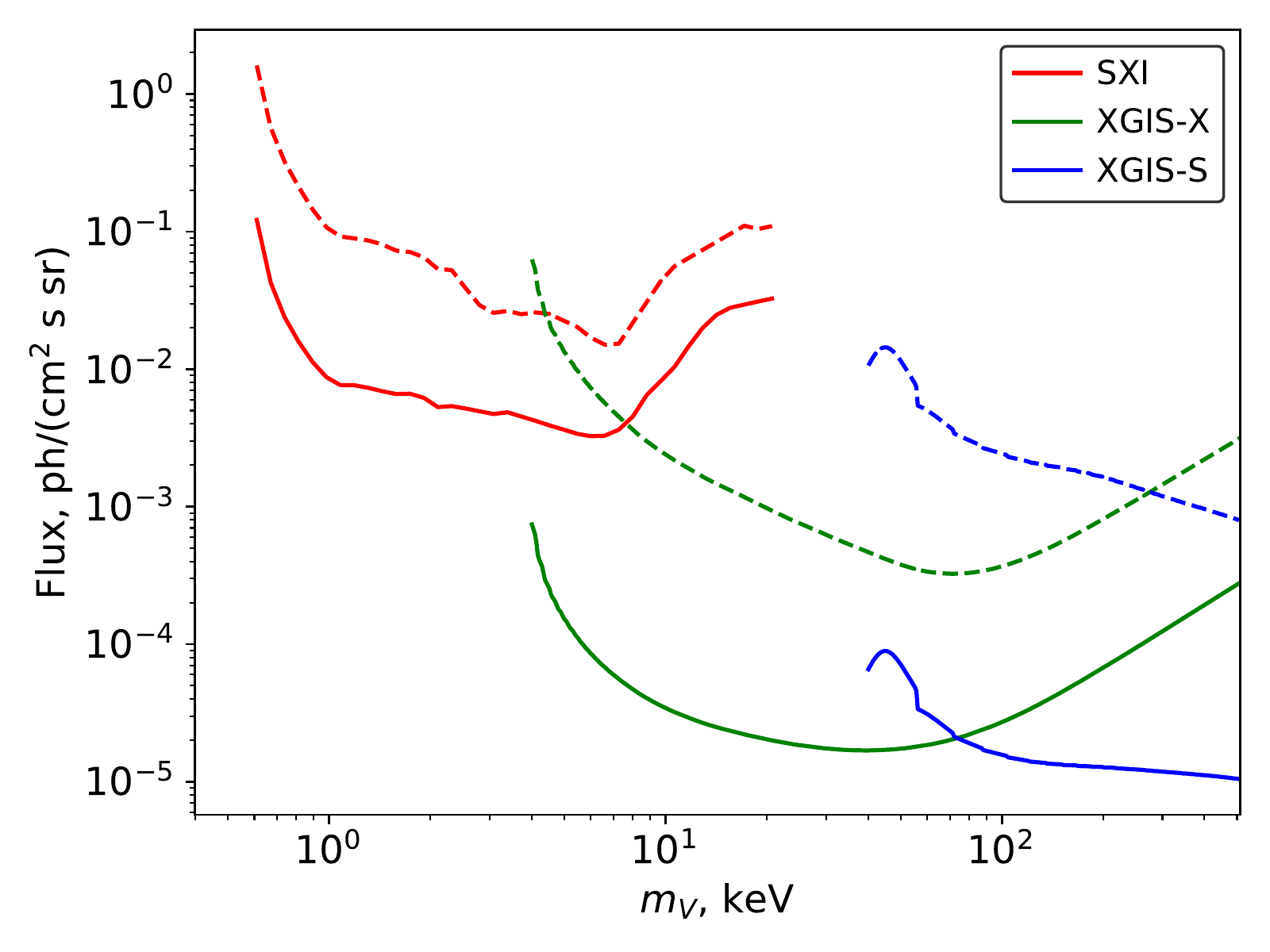}
   \caption{The $2\sigma$ limits on the sensitivity of \theseus' instruments to a narrow Gaussian line (\textit{left panel}) and a spectral feature expected for a dark photon decay (\textit{right panel}) (see Eq.~(\ref{eq:dm_flux}) from a 1~Msec long observation. The signal is assumed to be present in the whole FoV of the instrument. Dashed lines show the sensitivity of the instruments assuming a 1\% systematic error in each respectively. Results are presented as functions of the mass of the DM particle.}
    \label{fig:theseus_flux}
\end{figure*}

The DM-decay signal for each model respectively will be comprised of the spectrum given by Eq.~(\ref{eq:dm_flux}), and this signal is expected to be present in the real data on top of astrophysical and instrumental backgrounds. Such a signal can be distinguished from the background due to its characteristic shape (a narrow line for $\nu$MSM or ALPs; a relatively broad spectral feature in the case of dark photons). The minimal detectable flux for a given instrument depends on several factors and may be estimated as: 
\begin{align}
\label{eq:fmin}
& F_{min} = \sigma\left(\sqrt{\frac{B\Delta E}{A_{eff}T\Omega_{FoV}}}+\alpha B\Delta E\right)\quad \frac{\mbox{ph}}{\mbox{cm}^2\mbox{~s sr}}\,.  
\end{align}
Here, $T$ is the exposure of the observation (time duration for which data are taken), the instrument's effective area and energy resolution are denoted by $A_{eff}(E)$ and $\Delta E(E)$, respectively, and the observed background (instrumental and astrophysical) is $B(E)$~ph/(cm$^2$\,s\,keV\,sr). 
The parameter $\sigma$ stands for the significance level of the detection (e.g $\sigma=2$ for $2\sigma$ or $\sim 95$\% c.l.\ detection) and $\alpha$ for the level of characteristic systematic uncertainty of the instrument. We note that for DM candidates producing a signal that is broader than the instrument's energy resolution, one must utilise the characteristic width of the signal, instead of $\Delta E$.

Using Eq.~(\ref{eq:fmin}), the minimal detectable flux $F_{min}(E)$ derived from the data can be compared to the expected dark matter decay signal $F_{DM}(E)$ given by Eq.~(\ref{eq:dm_flux}) for each of the considered DM candidates. 
This allows the derivation of the range of dark matter parameters which the instrument is capable of probing.

\subsection{Observational strategy}
Any astrophysical object hosting a significant amount of dark matter can serve as a candidate for indirect searches for decaying dark matter. However, in order to maximise the potential of any instrument used, the object must have an angular size in the sky comparable to the instrument's FoV. Conversely, the observation of an object with a much smaller angular size than an instrument's FoV will suffer from a deterioration of the D-factor (and thus resulting DM decay flux), as the integral $D_{FoV}=\int D d\Omega $ vanishes beyond the characteristic size of the object. Therefore, neglecting to consider the relative size of an instrument's FoV can lead to the instrument's potential not fully being utilised. It is thus imperative to consider targets of a comparable angular size to the instrument's FoV. 

In the context of indirect DM detection, the suite of X-ray instruments aboard \theseus possess uniquely broad  fields of view ($\sim 1$~sr) which pose the issue of being larger than the angular size of any extra-galactic dark matter dominated object. Thus, with reference to the previous discussion of fields of view and object sizes, to fully utilise the capabilities of \theseus, we propose to focus on indirect DM searches of local, Milky Way concentrations of dark matter
with \theseus' instruments. Additionally, in order to minimise the level of astrophysical background (e.g. Galactic Ridge X-ray emission, GRXE) we propose that the observations should be located away from the galactic plane. Namely, we propose the observations to be located at latitudes $|b|>20$ where GRXE contributions are minimal~\cite{krivonos07}. We note, that a similar strategy was proposed for the eXTP and several other broad-FoV missions~\cite{we_extp,zhong20}.

To estimate the D-factor in the FoV of \theseus' instruments, and, consequently, the strengths of the expected decaying DM signals (see Eq.~(\ref{eq:dm_flux})), we assume that the density of the dark matter in the Milky Way follows a Navarro-Frenk-White (NFW~\cite{NFW_profile,nfw}) profile:
\begin{align}
\label{eq:nfw}
& \rho_{DM}(r) = \frac{\rho_0 r_0^{3}}{r(r+r_0)^2}\,, 
\end{align}
with $r_0\approx 17.2$~kpc and $\rho_0\approx 7.9\cdot 10^6\,M_{\bigodot}/\mbox{kpc}^3$ ~\cite{milky_way_profile}. 

The dark matter column density given by integrals in Eq.~(\ref{eq:D-factor}) was calculated numerically and the derived values of the D-factors for proposed for observations regions are summarized in Tab.~\ref{tab:observations}. We would like to note that the results presented below do not depend directly on the considered DM profile, but rather on the D-factor (dark matter column density) value. These results can be re-scaled according to the $D_{FoV}$ if another DM distribution model in the Milky Way is considered. We, however, expect the effect of re-scaling due to the variations in the dark matter profile to be relatively small. For example, an alternative profile considered in \cite{milky_way_profile} gives a D-factor of $\sim 10\%$ lower than the best fit considered model, consequently resulting in a marginally weaker signal.

To access the minimal detectable flux level (see Eq.~(\ref{eq:fmin})) within \theseus' observations, we estimated the expected  background flux level from simulated 1~Msec exposure blank sky observations with the \theseus/SXI and XGIS instruments. The simulated data were obtained with the \textit{fakeit} XSPEC (version: 12.10.1f) command, based on templates of blank sky observations provided by the \theseus collaboration\footnote{V7 templates dated May-July 2020; see \href{https://www.isdc.unige.ch/theseus/}{\theseus web page}} (\texttt{sxi\_bkg.pha}\footnote{Scaled by 17508, to account for template's FoV (675~arcmin$^2$).}, \texttt{XGIS-X\_0deg\_v7.bkg} and \texttt{XGIS-S\_0deg\_v7.bkg} ) and corresponding response files, see Fig.~\ref{fig:theseus_spec}. We note, that the provided templates are based on the estimations of both, instrumental and astrophysical backgrounds and thus any additional component(s) to model the background were not included.

The instrumental component of the backgrounds were simulated by the \theseus collaboration utilising the \texttt{GEANT4} toolkit. These backgrounds were  based on known low Earth orbit characteristic spectra and on the intensities of the particle background. The astrophysical background was adapted from the ROSAT All Sky Survey X-ray background, specifically from an area of sky of a  $10^\circ$-radius, centred on the North Ecliptic Pole \cite{x-ray_background}. 

We acknowledge that variations in the flux level between the simulations and observations will have an impact on the results of our simulations and therefore on our estimates of \theseus' capabilities. We argue that, a recent report on the background in the HXMT satellite has shown minimal variation between simulated (performed in the same manor as in \theseus) and observed backgrounds in an energy range similar to the SXI's~\citep{HXMT_background}. Namely, only a $5\% - 15\%$ increase is reported in the observed background, when compared to its counterpart in silico \cite{HXMT_background}. We argue that given the similar low Earth
orbits and energy ranges of HXMT and THESEUS, the simulations provide an accurate estimation of \theseus' background level.

Our simulations revealed that the spectral shapes of the background differ vastly between the SXI and XGIS-S/X instruments. The background of the SXI can be adequately modelled by the sum of two models representing both the astrophysical and instrumental backgrounds. The model of the astrophysical background was selected to be the sum of a power law and a hot thermal plasma with the temperature $\sim 0.2$~keV, constituting contributions from cosmic X-ray background and galactic X-ray emission~\cite{mccammon02,deluca04}. The instrumental background was best modelled by the sum of a power-law (not convolved with the effective area) and a set of four narrow Gaussian lines (see dot-dashed and dotted magenta lines in the left panel of Fig.~\ref{fig:theseus_spec} for corresponding model components).
For this instrument we therefore propose the use of the common observational strategy whereby one searches for a decaying-DM spectral feature on top of an adequately modelled background. This method is widely used in decaying dark matter searches in various astrophysical objects, see~\cite{sterile_neutrino_review19} for a review.

On the other hand, the backgrounds of the two XGIS detectors are characterised by a large number of line-like and broad spectral features (dominating the instrumental part of the background due to the coded-mask optics and SDD/CsI detectors in the instruments). We conclude that the XGIS' background is significantly more complicated than SXI's, and cannot be adequately modelled with any simple model. We thus propose the use of a different method for the XGIS, the ``ON-OFF'' observational strategy. This strategy requires the use of pairs of observations of a comparable duration, both ``ON'' and ``OFF'' the target. We propose to locate the ``ON'' observations closer to the Galactic Center than ``OFF'' ones, so $D_{FoV}^{ON}-D_{FoV}^{OFF}>0$. The estimations for $D_{FoV}^{ON}$ and $D_{FoV}^{OFF}$ for the sample ``ON'' and ``OFF'' observations are summarized in Tab.~\ref{tab:observations}. 

We acknowledge several  possible variations in the shape and intensities of the astrophysical and instrumental backgrounds including: energy, spatial and temporal fluctuations. These include variations in the instrumental/astrophysical background across the FoV of the instrument; variations in the instrumental background along the orbital path due to particle background variations, as well as variations in the astrophysical background between ``ON'' and ``OFF'' observation regions. To minimize the impact of orbital variations of the background we propose to perform ``ON'' and ``OFF'' observations in series of short consecutive observations, minimising the change in the above factors. Additional sources of background uncertainty include an imperfect modelling of the instrumental background and an imperfect knowledge of the instrument's response and effective area.

In the absence of detailed studies characterising the level of possible background variations for \theseus' instruments, we propose to estimate the impact of the previously described effects, by introducing a systematic uncertainty in \theseus' spectra. 
Below, we present all results for the case of an absence of systematic uncertainty, and compare them to the results in which a 1\% systematic uncertainty was introduced. Such a value of systematic uncertainty is characteristic for \xmm\footnote{See \href{https://xmmweb.esac.esa.int/docs/documents/CAL-TN-0018.pdf}{EPIC Calibration Status document} }. 
 In order to replicate the effects of systematic uncertainty, we introduced a new \texttt{STAT\_ERR} column to the simulated spectral files which, in addition to the standard Gaussian error, included a value proportional to the total counts in each channel.
 
 We note, that the systematics applied operate at small energy scales. Such   systematics effectively prevent uncertainties in each energy channel from being smaller than some fraction of the flux; independently of the observation's duration. Contrary to this, large scale systematics (applied to the whole spectrum or to its broad intervals) can change the overall normalisation of the spectrum (or its normalisation over broad intervals), without preventing the uncertainties in each interval becoming arbitrarily small. The characteristic level of large-scale systematics can be as large as $\sim 20-30$\% due to mis-modelling of the instrumental background~\citep[as in HXMT's case,][]{HXMT_background}; variability of the instrumental background along the orbital path due to particle background variations~\citep[see][for a discussion on \xmm background]{gastadello17} and  spatial variations of cosmic X-ray background~\citep{snowden97}. 
 
 For localised spectral feature searches large scale systematics, up to some level of accuracy, can be eliminated by the modification of certain features. Firstly one may alter the model considered for a fit (e.g. increase/decrease overall model normalisation in the simplest case); secondly, one may also modify the instrument's responses~\citep[see e.g. discussion in][]{cta_syst} or remedy this by splitting the data on a set of broad energy intervals and analysing each interval independently. Given otherwise arbitrarily small flux uncertainties (with increased observational time), these approaches can allow for the estimation of the impact of large-scale systematics on the flux of narrow features to be of the same order as the systematics, i.e. $\sim 20$\%. In what below we show that the 1\% small-scale systematic can imply substantially larger impact.

\subsection{Results}

Following the previously outlined methodology for simulating observations in both of \theseus{}' instruments, we conducted a search for a dark matter decay signal with a spectral shape (for each respective model) given by Eq.~(\ref{eq:dm_flux}), originating from the whole FoV.

We calculated $2\sigma$ upper limits on the normalisation of the signal, following Eq.~(\ref{eq:fmin}). For both the considered approaches (background modelling for the SXI and the ON-OFF technique for the XGIS) we adopted identical methods for the limit calculation. We note only that the only significant
difference between the two approaches was that
in terms of Eq.~(\ref{eq:fmin}) the background B  within the
characteristic signal width $\Delta E$ was estimated either from the model or from the ''OFF"-observation spectrum at the corresponding energy. The obtained limits allow us
 to derive the sensitivities of each of \theseus' instruments to the parameters of the DM particle in the corresponding model according to Eq.~(\ref{eq:dm_flux}).

The $2 \sigma$ ($\sim 95\%$ confidence level) limits on flux\footnote{Corresponding upper limits on the normalization were calculated with \texttt{error 4.0 } XSPEC command.} from 1~Msec long observation of Milky Way halo are shown with solid red, green and blue curves, for the SXI, XGIS-S and XGIS-X instruments respectively, in Fig.~\ref{fig:theseus_flux}. The left and right panels show the results for a narrow Gaussian line signal (sterile neutrino and ALP decay cases) and a broader spectral feature expected from a dark photon decay. Limits from observations where a 1\% systematic uncertainty was introduced to each instrument are shown with dashed lines.

The displayed limits illustrate that the sensitivity of each of \theseus' instruments to a DM decay signal is detrimentally affected by the effect of poorly controlled systematics for all of the types of DM particles considered. For a narrow line signal (sterile neutrino or ALP dark matter candidates) the SXI will suffer from worsening of its limits by a factor of $\sim 10$, whereas the XGIS is significantly more affected, seeing a reduction by a factor of $\sim 100$ in its sensitivity in both detectors. 
We therefore conclude that, despite the promising sensitivities of each instrument, instrumental systematics can be a significant obstacle and severely impair the ability of each instrument if not controlled.     
\begin{table}[t!]
\begin{tabular}{|c | c | c | c |}
\hline
Observation & FoV               & Galactic  &  $D_{FoV}$ \\ 
 &  deg$^2$   & coordinate centre&  GeV/cm$^2$ \\
\hline
SXI & $\sim 104^\circ\times 31^\circ$  & (110, 50) & $1\times10^{22}$~\\
Blank Sky & & & \\
\hline
XGIS & $\sim 104^\circ\times 31^\circ$ & On (0, 50)&  $2\times10^{22}$ \\
Blank Sky&  & Off (110, 50)& $1\times10^{22}$\\
\hline
\end{tabular}
\caption{Parameters of the simulated observations from blank sky readings. The FoV is assumed to be parallel to the galactic plane and roughly corresponds to the sky area at the border of which the effective area is 50\% of the on-axis one, see~\cite{XGIS}. Galactic coordinates show the coordinates of the FoV center in which the D-factor was calculated and the observation simulated (see text for details).}
\label{tab:observations}
\end{table}

\begin{figure*}[t!]
\hskip 1.cm ($\nu$MSM) \hskip 7.5cm (ALP)\\
\includegraphics[width=0.475\linewidth]{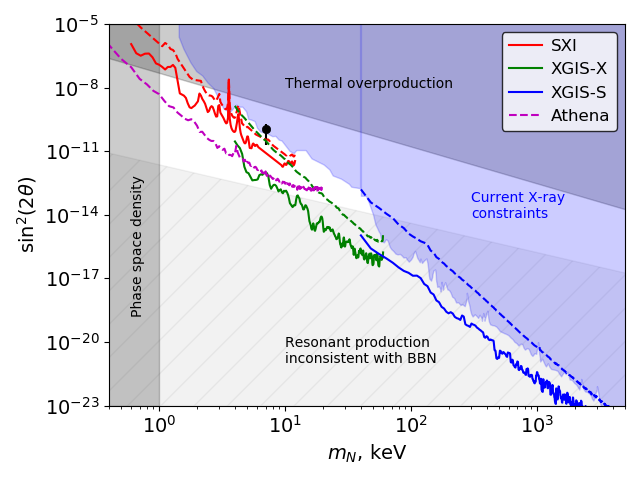}
\includegraphics[width=0.475\linewidth]{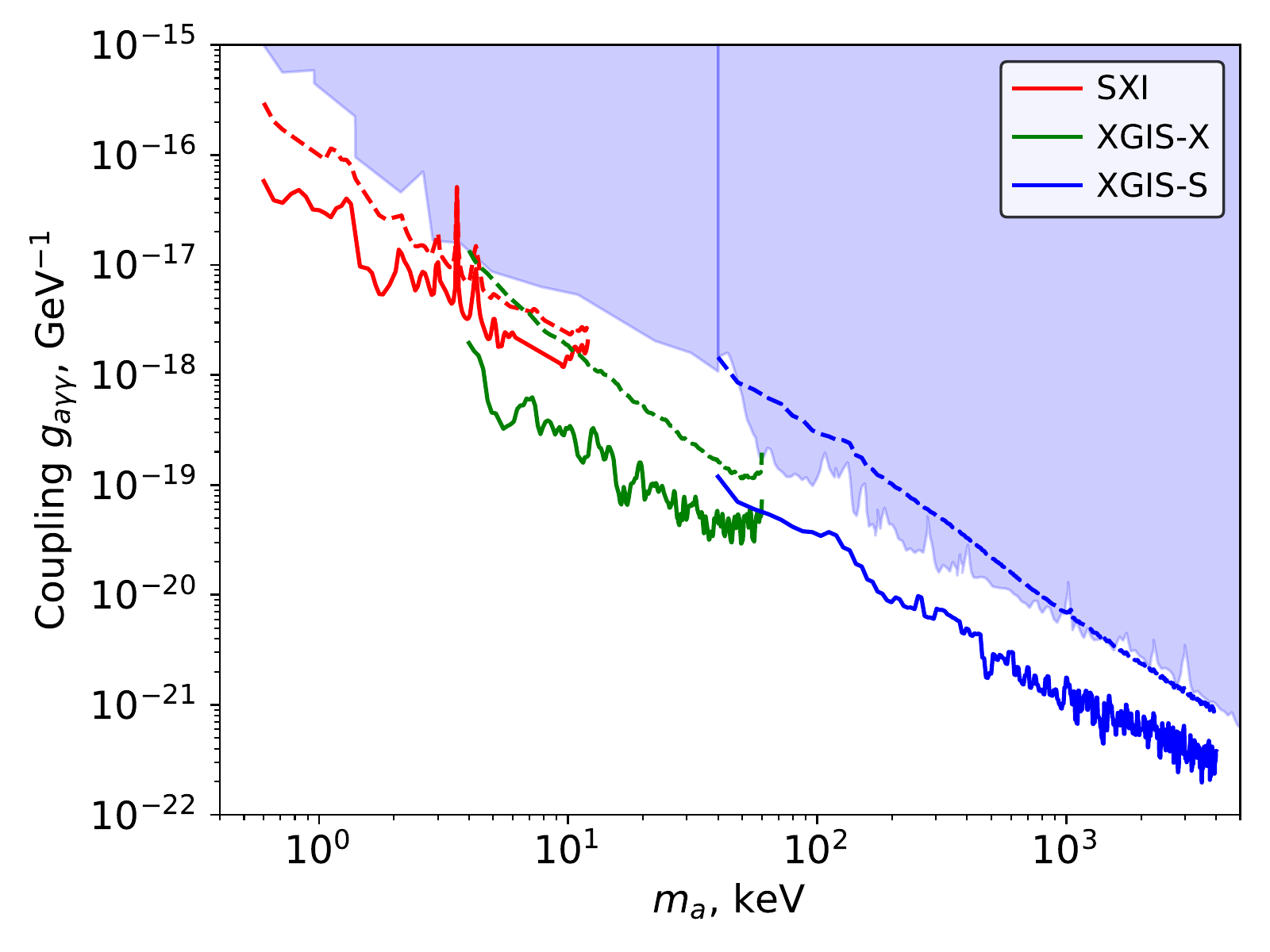}
\caption{The sensitivity of the instruments aboard \theseus{} to the parameters of sterile neutrino and ALP dark matter. All limits correspond to the $2\sigma$ values of the flux obtained from 1~Msec simulated observations of the Milky Way's DM halo, see Tab.~\ref{tab:observations} for the details. On all panels red, green and blue solid curves represent sensitivity limits for the SXI, XGIS-X and XGIS-S instruments, respectively, for a 0\% systematic uncertainty. Dashed curves illustrate similar limits for 1\% systematics present in the data. Shaded regions denote current exclusions adopted from~\cite{sterile_neutrino_review19,tremainegunn,tg1,gorbunov08,savchenko19,shi99,serpico05,shaposhnikov08,laine08,canetti13,pospelov08,graham13,fabbrichesi20, we_spi,redondo13}. \textit{Left panel}: \theseus' sensitivity for the sterile neutrino ($\nu$MSM) DM. The Magenta curve illustrates limits reachable for Athena~\cite{athena_limits}. The black point represents the parameter point corresponding to the tentative detection of an $\sim 3.55$~keV line in certain DM-dominated objects~(see \cite{line35},\cite{line35_1} and \cite{sterile_neutrino_review19} for a recent review). \textit{Right panel}: Sensitivity limits for ALP dark matter, see text for details.}
\label{fig:dm_limits}
\end{figure*}

\begin{figure}[ht!]
\hskip 1cm (Dark photon)\\
\includegraphics[width=0.95\linewidth]{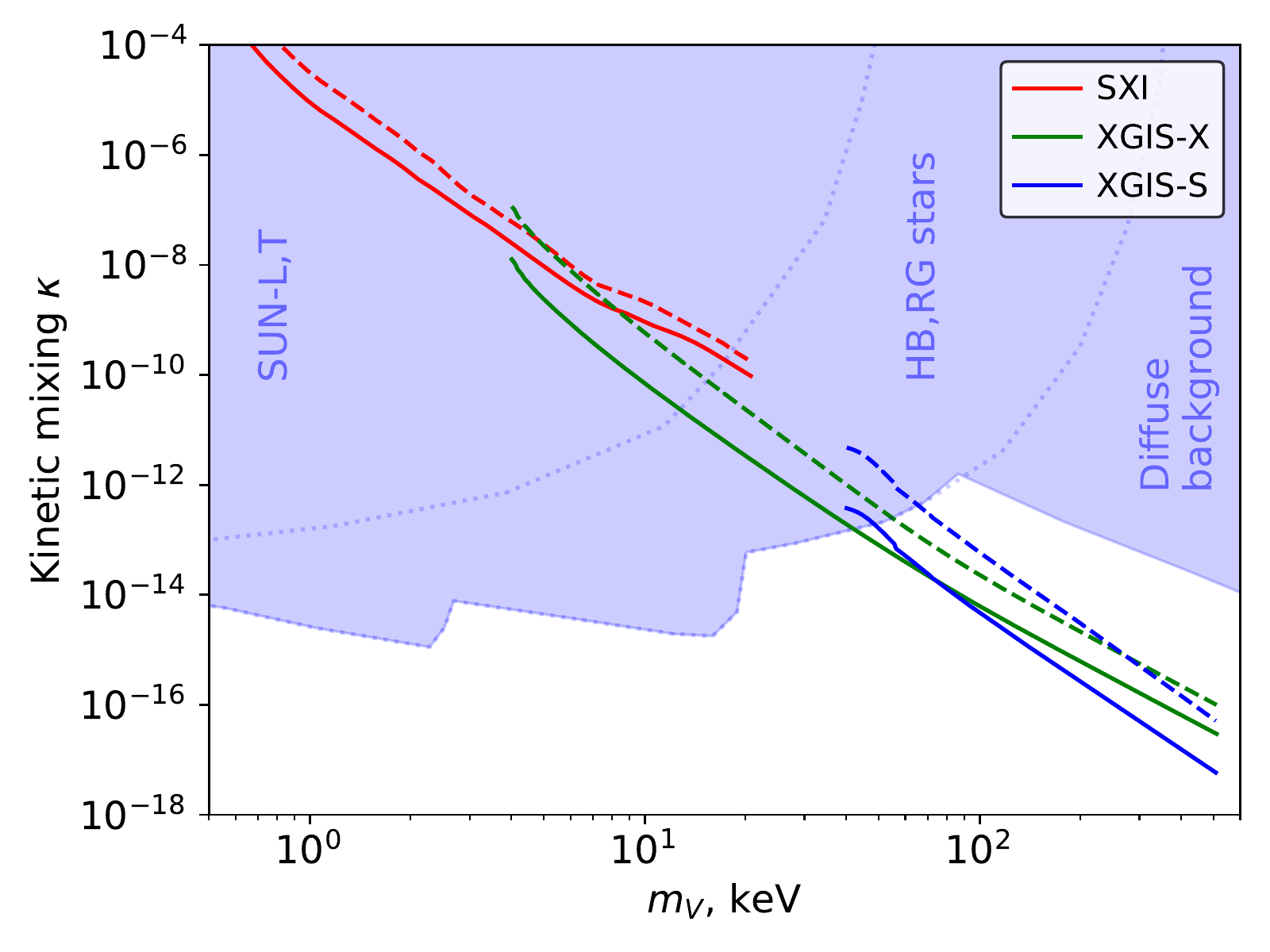}
\caption{The sensitivity of the instruments aboard \theseus{} to the parameters of dark photon dark matter, see caption of Fig.~\ref{fig:dm_limits} for designations and text for details.}
\label{fig:dm_limits1}
\end{figure}

For each of the considered DM models (ALPs, sterile neutrino and dark photons), we converted the obtained flux limits into limits on the parameters of the DM particles, see Eq.~(\ref{eq:dm_flux}) and Figs.~\ref{fig:dm_limits}-\ref{fig:dm_limits1}, and compared the obtained limits to other limits presented in the literature.

For the sterile neutrino we compared limits derived by this study to  existing observational X-ray and $\gamma$-ray constraints (see \cite{sterile_neutrino_review19} for a review). We also display, for comparison, the expected limits from 1~Msec-long Segue~I dSph observations by the forthcoming \textit{Athena} mission~\cite{athena_limits}, given a zero level of systematic uncertainty. The limits based on the phase space density arguments for the DM in dSphs~\cite{tremainegunn,tg1,gorbunov08,savchenko19} and otherwise incorrect abundance of sterile neutrinos produced in the Early Universe~\cite{dodelson,numsm1} are shown as gray shaded regions. Model dependent limits based on parameter values that are inconsistent with the observed abundances of light elements produced during Big Bang Nucleosynthesis~\cite{shi99,serpico05,shaposhnikov08,laine08,canetti13} (see, however,~\cite{kusenko06}) are shown as a gray hatched region.

The limits on ALPs were compared to the existing limits in the keV band based on the non-detection of a line-like feature in the spectrum of diffuse gamma-ray background in the keV-MeV band~\cite{graham13,we_spi}. The limits on dark photons, on the other hand, are compared to stellar evolution-based limits (the Sun limits in longitudinal and transverse channels; the limits from horizontal branch and red giant stars' evolution~\cite{redondo13}) and the limits from the diffuse gamma-ray background, see \cite{pospelov08,fabbrichesi20} for a review.

\section{Discussion and Conclusions}
This study has investigated the sensitivity of the proposed X-ray telescope arrays aboard the upcoming \theseus{} mission to decaying dark matter signals from DM models with ALPs, sterile neutrinos and dark photons. Our results demonstrate that \theseus{} has the potential to impose significantly better limits than the current generation of instruments. The use of 1~Msec long \theseus{} observations of blank sky regions has the potential to improve existing X-ray constraints on the parameters of dark matter, by a factor of up to $\sim 300$, within the keV-MeV dark matter particle mass range, see Fig.~\ref{fig:dm_limits} and Fig.~\ref{fig:dm_limits1}.  

The regions proposed for observations are located at significant angular distance from the Galactic Center. This allows the minimisation of uncertainties connected to the knowledge of the exact shape of the dark matter profile and excludes the presence of strong astrophysical backgrounds. In case the \theseus mission is approved with reduced specifications\footnote{See recent updates on \href{https://www.isdc.unige
.ch/theseus/mission-payload-and-profile.html}{\theseus mission website}}, the relocation of observational regions closer to the Galactic Center can compensate (within a factor of $\sim 2$) the subsequent decrease in the expected dark matter signal.

We also show that the XGIS has the potential to completely explore the sterile neutrino parameter space in the mass range $m_N\sim 15 - 150$ keV  (see Fig.~\ref{fig:dm_limits}, left panel), assuming a marginally possible $0\%$ level of systematic uncertainty.

We assert that the effect of systematics on \theseus' instruments will be severely detrimental to their sensitivity to all types of decaying DM. We have shown a level of systematics at $1\%$ can considerably worsen the constraints that can be achieved by both instruments, with the limits imposed by the SXI and XGIS falling by up to factors of $\sim 10$ and $\sim 100$ respectively for all considered DM models. At these levels of systematic uncertainty, while the XGIS will remain able to probe new areas of the parameter space, the SXI's limits may, in certain ranges, be worse than the existing limits in this energy band. To summarise, only  full control of the systematics in these instruments would make them a formidable addition in the search for DM.

The tentative detection of a $3.55$~keV line in some DM-dominated objects~\cite{line35_1, line35} is still actively being discussed in the field (see~\cite{sterile_neutrino_review19} for a recent review). Such a signal was originally proposed to originate from the decay of a sterile neutrino with the mass $m_N\sim 7$~keV and a mixing angle of ($\sin^2(2\theta)\sim 2\cdot 10^{-11}$). The corresponding range of mixing angles discussed in the literature is denoted by the black point with error-bars in Fig.~\ref{fig:dm_limits}. We mention that the  constraints displayed in Fig.~\ref{fig:dm_limits} for a $0\%$ systematic uncertainty (left panel) indicate also that \theseus{} will be sensitive enough to exclude or detect this line, at a $\gtrsim 7\sigma$ level ($\sim 3\sigma$ level if 1\% systematics is present).  The strength of such a line could be compared to other DM-dominated objects or along the sky in order to correlate its intensity with the known $D_{FoV}$ value, and thus draw conclusions on its possible DM-decay origin.

We would further like to note that several other models were proposed to explain the observed 3.55~keV signal. These models include scalar~\cite{babu14} and pseudo-scalar, ALP~\cite{alps355,alps_355_1} dark matter. We argue that the (non)detection of such a line or feature by \theseus can provide significant constraints on the parameters of these models.

The \theseus{} mission, as well as its numerous scientific objectives, will play an essential part in high energy studies over the next decade. Its overlap with other planned missions such as eXTP and Athena provides prime potential for the complementary study of the decaying DM's parameter space using the above mentioned next generation satellites, among many others. The use of these instruments in conjunction with one-another has the potential to impose tighter limits on DM candidates than ever before and significantly decrease their unexplored parameter space.

We conclude that \theseus{}, alongside well controlled systematics, has the potential to either detect decaying dark matter, or to impose some of the strongest constraints on its properties among its generation of satellites. 
\\
\\
\noindent\textit{Acknowledgements}
The authors acknowledge support by the state of Baden-W\"urttemberg through bwHPC. This work was supported by DFG through the grant MA 7807/2-1.

\def\aj{AJ}%
\def\actaa{Acta Astron.}%
\def\araa{ARA\&A}%
\def\apj{ApJ}%
\def\apjl{ApJ}%
\def\apjs{ApJS}%
\def\ao{Appl.~Opt.}%
\def\apss{Ap\&SS}%
\def\aap{A\&A}%
\def\aapr{A\&A~Rev.}%
\def\aaps{A\&AS}%
\def\azh{AZh}%
\def\baas{BAAS}%
\def\bac{Bull. astr. Inst. Czechosl.}%
\def\caa{Chinese Astron. Astrophys.}%
\def\cjaa{Chinese J. Astron. Astrophys.}%
\def\icarus{Icarus}%
\def\jcap{J. Cosmology Astropart. Phys.}%
\def\jrasc{JRASC}%
\def\mnras{MNRAS}%
\def\memras{MmRAS}%
\def\na{New A}%
\def\nar{New A Rev.}%
\def\pasa{PASA}%
\def\pra{Phys.~Rev.~A}%
\def\prb{Phys.~Rev.~B}%
\def\prc{Phys.~Rev.~C}%
\def\prd{Phys.~Rev.~D}%
\def\pre{Phys.~Rev.~E}%
\def\prl{Phys.~Rev.~Lett.}%
\def\pasp{PASP}%
\def\pasj{PASJ}%
\def\qjras{QJRAS}%
\def\rmxaa{Rev. Mexicana Astron. Astrofis.}%
\def\skytel{S\&T}%
\def\solphys{Sol.~Phys.}%
\def\sovast{Soviet~Ast.}%
\def\ssr{Space~Sci.~Rev.}%
\def\zap{ZAp}%
\def\nat{Nature}%
\def\iaucirc{IAU~Circ.}%
\def\aplett{Astrophys.~Lett.}%
\def\apspr{Astrophys.~Space~Phys.~Res.}%
\def\bain{Bull.~Astron.~Inst.~Netherlands}%
\def\fcp{Fund.~Cosmic~Phys.}%
\def\gca{Geochim.~Cosmochim.~Acta}%
\def\grl{Geophys.~Res.~Lett.}%
\def\jcp{J.~Chem.~Phys.}%
\def\jgr{J.~Geophys.~Res.}%
\def\jqsrt{J.~Quant.~Spec.~Radiat.~Transf.}%
\def\memsai{Mem.~Soc.~Astron.~Italiana}%
\def\nphysa{Nucl.~Phys.~A}%
\def\physrep{Phys.~Rep.}%
\def\physscr{Phys.~Scr}%
\def\planss{Planet.~Space~Sci.}%
\def\procspie{Proc.~SPIE}%
\let\astap=\aap
\let\apjlett=\apjl
\let\apjsupp=\apjs
\let\applopt=\ao
\bibliography{biblio.bib}

\end{document}